# Imaging the Lyman-alpha Forest

**Andrew Gould**[1]

and

**David H. Weinberg**

Dept of Astronomy, Ohio State University, Columbus, OH 43210

gould@payne.mps.ohio-state.edu, dhw@payne.mps.ohio-state.edu

**Abstract**

We show that it is now possible to image optically thick Ly$\alpha$ clouds in fluorescent Ly$\alpha$ emission with a relatively long ($\sim 20\,\mathrm{hr}$) integration on a large ($\sim 10\,\mathrm{m}$) telescope. For a broad range of column densities ($N \gtrsim 10^{18.5}\,\mathrm{cm}^{-2}$), the flux of Ly$\alpha$ photons from recombination cascades is equal to $\sim 0.6$ times the flux of ionizing photons, independent of the geometry of the cloud. Additional Ly$\alpha$ photons are produced by collisional excitations when these are the cloud's primary cooling mechanism. For typical physical conditions expected in optically thick clouds, these mechanisms together lead to a Ly$\alpha$ emission flux that is $\sim (2/3)\langle\nu\rangle/\nu_0$ times the flux of ionizing photons, where $\langle\nu\rangle$ is the mean frequency of ionizing background photons and $\nu_0$ is the Lyman limit frequency. Hence, measurement of the surface brightness from an optically thick cloud (known to exist, e.g., from a quasar absorption line) gives a direct measure of the energy in the ionizing radiation background. Moreover, in the same long slit spectrum one could hope to detect emission from $\sim 200$ other Ly$\alpha$ systems. Such detections would allow one to make a 2-dimensional map of the Ly$\alpha$ forest. By taking a series of such spectra, one could map the forest in three dimensions, revealing structure in the high-redshift universe.

---

1 Alfred P. Sloan Foundation Fellow

## 1. Introduction

Resonant absorption against the continuum of a bright background source is an exquisitely sensitive way to detect diffuse gas. The Ly$\alpha$ forest in the spectra of high redshift quasars provides a detailed picture of structure in the universe at $z \sim 2-5$, revealing systems that are much more common than galaxies but whose internal gas densities are far lower. Each quasar spectrum yields a 1-dimensional map of the neutral hydrogen distribution along its line of sight. Our limited information about the sizes and morphologies of Ly$\alpha$ absorbers comes from the rare instances where two nearby quasars (or the multiple images of a gravitational lens) probe neighboring lines of sight. In this paper, we argue that 10-m class telescopes make it feasible, though still challenging, to detect the optically thick end of the Ly$\alpha$ forest (neutral column density $N \gtrsim 10^{18}\mathrm{cm}^{-2}$) in Ly$\alpha$ emission, thereby building up a 3-dimensional picture of the properties and space distribution of these systems.

There have been many attempts to detect Ly$\alpha$ emission from high redshift "primeval galaxies." Some of these efforts have covered large areas of sky (e.g., Thompson, Djorgovski, & Trauger 1995; Thompson & Djorgovski 1995), while others have targeted known damped Ly$\alpha$ absorbers (e.g., Lowenthal et al. 1995), which are plausible progenitors of disk galaxies. The motivating idea behind these searches is that regions of active star formation should produce large numbers of Ly$\alpha$ photons. The paucity of successful detections has therefore been interpreted as evidence for either low rates of star formation or internal dust extinction in these systems. However, hydrogen clouds should emit some Ly$\alpha$ radiation even if they have no ongoing star formation at all, because the metagalactic UV background photoionizes the clouds (or at least their optically thin skins), and many of the resulting recombinations produce Ly$\alpha$ photons that ultimately escape.

The idea of searching for this fluorescent emission from Ly$\alpha$ systems has been discussed previously, e.g. by Hogan & Weymann (1987) and by Lowenthal et al. (1990), who used their null result to place an upper limit on the UV background intensity. Hogan & Weymann (1987) estimated the surface brightness of optically thin ($N \lesssim 10^{17}\mathrm{cm}^{-2}$) clouds and concluded that these might plausibly be detected with 2-m class telescopes. Their conclusions were substantially more optimistic than those reported below, largely because they assumed a fiducial UV background intensity more than 100 times higher than the more recent observational estimates that we adopt. With the lower background intensity, optically thin clouds are difficult to detect even with 10-m telescopes. We therefore emphasize the possibility of detecting optically thick systems, which, roughly speaking, absorb the UV continuum radiation and mirror it back into space in the concentrated form of a double-peaked Ly$\alpha$ line. We discuss a number of physical effects that influence the optimal design of search strategies and the interpretation of detections and null results.

Ionizing radiation gives rise to Ly$\alpha$ emission in two distinct ways. First, in ionization equilibrium, each ionization leads to a recombination, and most of these in turn ultimately yield a Ly$\alpha$ photon. Second, in thermal equilibrium, the kinetic energy of the ionized electron heats the gas, and this heat is ultimately emitted in radiation. Most of this radiation is in the form of free-free emission, free-bound emission, or line emission induced by collisional excitation. At high densities and temperatures $T \sim 2 - 5 \times 10^4$ K, conditions that are likely to prevail in optically thick clouds, line emission is the dominant cooling mechanism, so a large fraction of the deposited thermal energy also comes out in Ly$\alpha$ photons.

Our approach in this paper is first to analyze the recombination Ly$\alpha$ photon production, which depends primarily on the total *number* of ionizing photons. We give a simple analytic treatment in § 2, then present the results of numerical simulations in § 3, where we focus on the frequency and angular distribution of the emitted line photons. In § 4 we analyze the effects of cloud topology (coherent absorbers vs. clusters of cloudlets) on the expected form of the emission signal. In § 5, we turn to the excitation Ly$\alpha$ photon production, which depends primarily on the total *energy* of the ionizing photons. We show that, for optically thick clouds, the total Ly$\alpha$ emission is proportional to the total energy of the ionizing background. Finally, in § 6, we present a signal-to-noise estimate for a fiducial observation and discuss what could be learned from various types of searches for emission from Ly$\alpha$ systems.

## 2. Recombination Emission from Lyman $\alpha$ Clouds

Consider first a Ly$\alpha$ cloud which is optically thin to ionizing radiation, $N \ll 10^{17}\,\mathrm{cm}^{-2}$. The hydrogen atoms will then be ionized at a rate $\Gamma$,

$$\Gamma = 4\pi \int_{\nu_0}^{\infty} d\nu \phi(\nu)\sigma(\nu), \qquad \phi(\nu) \equiv \frac{J(\nu)}{h\nu}, \tag{2.1}$$

where $J(\nu)$ is the photon flux density (in $\mathrm{erg\,cm^{-2}\,s^{-1}\,sr^{-1}\,Hz^{-1}}$), $\sigma(\nu)$ is the ionization cross section, and $h\nu_0 = 13.6\,\mathrm{eV}$ is the Rydberg energy. We assume that the cloud is in ionization equilibrium, i.e., the ionization rate equals the recombination rate, which should be a good approximation for Ly$\alpha$ clouds. A typical recombination leads ultimately to the production of a single Ly$\alpha$ photon, and this photon resonantly scatters off H atoms until it escapes the cloud. There are, however, several loss mechanisms. First, for a broad range of relevant temperatures centered on $T = 10{,}000$ K, $\sim 38\%$ of recombinations go directly to the ground state (Osterbrock 1989). These recombination photons escape an optically thin cloud without generating Ly$\alpha$ photons. Second, a fraction $\sim 32\%$ (which is again only weakly dependent on $T$) of recombinations ultimately produce excited H in the $2S$ rather than $2P$ state (Osterbrock 1989). These atoms decay by 2-photon rather than Ly$\alpha$ emission. Hence a fraction $\eta_{\rm thin} \sim 42\%$ of recombinations ultimately yield a Ly$\alpha$ photon. If the cloud has area $l^2$, a cosmologically nearby observer ($z \ll 1$) at distance $D$ would see a flux of Ly$\alpha$ photons $\eta_{\rm thin} N l^2 \Gamma / 4\pi D^2$, corresponding to a number flux (photons $\mathrm{cm^{-2}\,s^{-1}\,sr^{-1}}$) of $\Phi_{\rm obs} = \eta_{\rm thin} N \Gamma / 4\pi$. At cosmological distances, bolometric surface brightness is reduced by $(1+z)^{-4}$, so photon number density is reduced by $(1+z)^{-3}$. Hence the observable Ly$\alpha$ photon flux is,

$$\Phi_{\rm obs} = (1+z)^{-3}\eta_{\rm thin} N \int_{\nu_0}^{\infty} d\nu \phi(\nu)\sigma(\nu) \qquad \text{(optically thin)}, \tag{2.2}$$

where $\eta_{\rm thin} \sim 42\%$. The cross section for hydrogen ionization is (Osterbrock 1989),

$$\sigma_{\rm H}(\nu) = \frac{\sigma_0}{1-\exp(-2\pi/\epsilon)}\left[\frac{\nu_0}{\nu}\exp\left(1-\frac{\tan^{-1}\epsilon}{\epsilon}\right)\right]^4 \approx \sigma_0\left(\frac{\nu}{\nu_0}\right)^{-2.75}, \tag{2.3}$$

where

$$\sigma_0 = 6.3\times 10^{-18}{\rm cm}^2, \qquad \epsilon \equiv \left(\frac{\nu}{\nu_0}-1\right)^{1/2}. \tag{2.4}$$

Consider next a spherical cloud of radius $r$ which is optically thick to ionizing radiation. Every incident photon with $\nu > \nu_0$ will ionize a hydrogen atom, which will then recombine and, in most cases, produce a Ly$\alpha$ photon. It is still true that $\sim 38\%$ are captured directly into the ground state, but most of the resulting continuum photons are captured again. We find numerically in § 3 that only $\sim 5\%$ leave the cloud without converting into a line photon. As in the optically thin case, $\sim 32\%$ of line photons end in 2-photon decays. In § 3, we estimate the fraction of the remaining Ly$\alpha$ photons which are absorbed by dust before they escape. For our fiducial cloud with $N \sim 10^{19}\,{\rm cm}^{-2}$, this fraction is $\sim 4\%$. The cloud will therefore emit Ly$\alpha$ photons at a rate $\eta_{\rm thick} \sim 62\%$ times the rate at which it absorbs ionizing radiation, i.e., $4\pi^2\eta_{\rm thick}r^2 \int d\nu\phi(\nu)$. (The parameter $\eta_{\rm thick}$ has a weak dependence on HI column density, as shown in § 3, below.) We now assume that the cloud has a "sharp" boundary, so that the region of continuum absorption is the same as the region of Ly$\alpha$ emission. Making use of this rather strong assumption, whose validity we discuss in § 3, we repeat the above calculation and find,

$$\Phi_{\rm obs} = (1+z)^{-3}\eta_{\rm thick} \int_{\nu_0}^{\infty} d\nu \phi(\nu) \qquad \text{(optically thick)}. \tag{2.5}$$

While this calculation was performed for a spherical cloud, it actually remains valid for any sharp-edged optically thick cloud. This conclusion follows from detailed balance and the fact that both the ambient radiation and the emission process are (by assumption) isotropic. Note that we have ignored ionization of HeII by energetic photons ($\nu > 4\nu_0$). We include HeII ionizations in our numerical simulations reported below, but their net effect is small.

## 3. Ly$\alpha$ Scattering

In an optically thick cloud, ionizing photons will be absorbed at an (ionizing) optical depth $\sim 1$, corresponding to column depth $N \sim (\nu/\nu_0)^{2.75}\sigma_0^{-1} \sim 10^{18}\,\mathrm{cm}^{-2}$, where we have taken $\nu/\nu_0 \sim 2$. This corresponds to an optical depth $\tau \sim 10^4$ for the emitted Ly$\alpha$ photons, if the cloud's 1-dimensional, rest-frame velocity dispersion (which may reflect both thermal broadening and bulk internal motions) is $\sigma \sim 35\,\mathrm{km\,s}^{-1}$. Scattering of Ly$\alpha$ photons in an optically thick hydrogen cloud has been studied for at least half a century (Holstein 1947; Neufeld 1991 and references therein). However, the geometry of the present problem differs significantly from that considered in most previous treatments. In particular, our Ly$\alpha$ photons are generated in the cloud's "shielding" layer, and since this is likely to be small compared to the cloud, the appropriate geometry is a slab. We therefore focus on the problem of the distribution in redshift and angle of photons which escape from such a slab. The bottom line is rather simple: a typical Ly$\alpha$ photons "scatters" (more precisely, it is absorbed and reemitted) until it is emitted by an atom with velocity $\sim \pm\sqrt{2\ln(\tau)}\sigma \sim \pm 4\sigma$ relative to the central velocity of the cloud. The emission is isotropic, and the profile is double peaked at the $\sim 4\,\sigma$ wings of the cloud.

In the physical conditions of Ly$\alpha$ clouds, the time between photon absorption and emission is very short compared to the collision time and can be neglected. Hence, the absorption and emission together can be regarded as scattering. The scattering is not quite isotropic because of the $\sin^2\theta$ dependence of the $1S$–$2P$ transition, but this anisotropy has virtually no effect. As we show below, the typical column densities at escape are $N \sim 10^{18}\,\mathrm{cm}^2$, so the damping wings of the Ly$\alpha$ line are unimportant. Hence, the photon is always absorbed by an atom with the same velocity projected along its direction of motion as that of the atom which emitted it.

Although the escape of Ly$\alpha$ photons from a cloud is often called "diffusion", this process is *not* primarily diffusive in either physical or frequency space. Rather, the photons tend to maintain their initial spatial distribution, and up to one or two scatters before a photon finally escapes, its redshift is near the mean of the cloud. Each photon in this distribution suffers a single catastrophic scatter which suddenly moves its dimensionless Doppler shift $\Delta$ to the extreme tail of the distribution, $\Delta \sim \pm 4$. Here,

$$\Delta \equiv \frac{\delta\nu/\nu}{\sigma/c}, \tag{3.1}$$

where $\delta\nu$ is the Doppler shift, $\nu$ is the line center, and $\sigma$ is the 1-dimensional dispersion of the gas. The photon then faces a very low effective optical depth $\tau_{\rm eff} = \tau \exp(-\Delta^2/2)$ where $\tau$ is the optical depth at line center. It therefore escapes directly from high $\tau$. This simple model of the escape process, justified analytically and numerically below, allows one to make an accurate estimate of the distribution of emitted photons as a function of redshift and angle. In brief, for an initial distribution at characteristic optical depth $\tau_0$, the photons are emitted with angular dependence $\propto \cos\theta\, d\cos\theta$ (just as is the case for diffusive escape) and with redshift sharply peaked at $\pm\Delta_{\rm peak}$, where $\Delta_{\rm peak} \sim \sqrt{2\ln(\tau_0 \sec\theta)}$ (which is substantially different from diffusive escape). The width of each of these peaks is $\sim \Delta_{\rm peak}^{-1}$. Here $\theta$ is the angle of escape relative to the normal of the surface of the cloud, which is assumed to have little curvature on scales of the physical depth at which the photons are captured.

After any given scatter, the probability that the photon is shifted by $\Delta$ is $p(\Delta) = (2\pi)^{-1/2}\epsilon(\Delta)$, where $\epsilon(\Delta) \equiv \exp(-\Delta^2/2)$. Such photons will encounter a reduced density of hydrogen scatterers and will therefore travel a factor $\epsilon^{-1}$ farther than photons at the line center before colliding again. [The density of wing photons in the gas at any given time is therefore uniform in $\Delta$, i.e., with no Boltzmann factor, $\exp(-\Delta^2/2)$.] If one of these photons is headed in the direction $\theta$, then it will face an effective optical depth $\tau_{\rm eff} = \epsilon\tau\sec\theta$ and so will escape with probability $\exp(-\tau_{\rm eff})$. Hence, at fixed optical depth $\tau$, the rate of escape $\Gamma$ is,

$$\frac{d\Gamma}{d\cos\theta d\Delta} \propto \epsilon e^{-\epsilon\tau\sec\theta}. \tag{3.2}$$

At fixed $\theta$, the peak of this distribution is at

$$\Delta_{\rm peak}(\theta) = \sqrt{2\ln(\tau\sec\theta)}, \tag{3.3}$$

and the integrated emission rate,

$$\frac{d\Gamma}{d\cos\theta} \propto \frac{\cos\theta}{\tau\Delta_{\rm peak}}, \tag{3.4}$$

is very nearly proportional to $\cos\theta$. The width of each of the two peaks is $\sim \Delta_{\rm peak}^{-1}$.

The distribution of optical depths of the Ly$\alpha$ photons prior to escape can be roughly characterized as an exponential, $\exp(-\tau/\tau_0)d(\tau/\tau_0)$. We focus on the emission at fixed angle, $\theta$. Since the distribution in $\Delta$ at fixed $\tau$ is comprised of two narrow peaks, we approximate it as two $\delta$-functions at $\Delta = \pm\Delta_{\rm peak}$. From equation (3.3) and the definition $\epsilon = \exp(-\Delta^2/2)$ we find that $\tau = \epsilon^{-1}\cos\theta$, and this substitution leads to a total emission rate,

$$\frac{d\Gamma_{\rm tot}}{d\cos\theta d\Delta} \propto \Delta\epsilon^{-1}e^{-\cos\theta/\epsilon\tau_0}. \tag{3.5}$$

This distribution is narrowly peaked around $\Delta \sim \pm\sqrt{2\ln(\tau_0\sec\theta)}$. The net result is that ionizing photons penetrate to a typical (Ly$\alpha$) optical depth $\tau \sim 10^4$, then the emission takes place from a similar depth, with an angular distribution very similar to that from a diffusive optically thick surface, and with a shift distribution composed of two sharp peaks at $\Delta \sim \pm 4$.

We confirm this picture by means of a Monte Carlo simulation. We first irradiate an optically thick slab of hydrogen ($N = 10^{19}\,\mathrm{cm}^{-2}$) with an ionizing spectrum taken from Haardt & Madau (1995; hereafter HM). HM compute the ambient UV background that would be produced by the emission from observed quasars after it is filtered through the Ly$\alpha$ forest; for the calculations in this paper, we use HM's $z = 2$ spectrum. (We assume that for $\nu > 4\nu_0$, the ionization cross section is $12.5\sigma_{\mathrm{H}}(\nu/4)$, corresponding to a ratio of HeII to HI of 50. However, this assumption has relatively little effect because the HM spectrum has very few such photons.) We determine the fraction of photons which pass through the cloud without interacting (0.5%) and the fraction which generate Lyman continuum photons that escape without forming line photons (5%). Next, we allow each resulting Ly$\alpha$ photon to scatter until it leaves the cloud, and we tabulate $N_{\mathrm{walk}}$, the total column density through which the photon randomly walks in the course of escaping. We then compute the probability that each photon is absorbed by dust, $1 - \exp(-N_{\mathrm{walk}}/N_*)$, where $N_* = 10^{21}\,\mathrm{cm}^{-2}$. This value is adopted based on the work of Fall & Pei (1995), who argue that damped Ly$\alpha$ systems of column density $N_*$ typically extinguish about 1 mag of background quasar light. We find that dust absorption is small, $\sim 4\%$. Of course, the absorption could be still lower if the dust-to-gas ratio for $N \sim 10^{19}\,\mathrm{cm}^{-2}$ systems is lower than for damped systems.

Figure 1 shows the cumulative angular distribution of the emitted photons compared to the curve ($\cos^2\theta$) one would expect for optically thick diffusive emission. There is virtually no difference.

Figure 2 shows the dimensionless Doppler shift $\Delta$ summed over all emission angles. As expected there are two narrow peaks. We have confirmed (but not shown in the Figure) that the distributions at fixed emission angle have even narrower peaks, whose separations grow $\propto \sqrt{\ln(\tau \sec\theta)}$.

Figure 3 illustrates the essentially non-diffusive nature of the emission process. The solid and solid bold histograms show the initial and pre-emission distributions of the photons in optical depth. The two distributions are broadly similar. It is true that the photons are typically emitted at about half the optical depth at which they are absorbed, but since the characteristics of emission are determined by $\sqrt{\ln(\tau)}$, this reduction does not qualitatively alter our basic picture. The small role played by diffusion can be understood as follows. The distribution of Doppler shifts is $\sim \epsilon(\Delta)$. Typically a photon will directly escape after $N \sim \tau$ collisions by being scattered to $\Delta_{\rm peak} \sim \sqrt{2\ln\tau}$. We can estimate the number of mean free paths (mfps) traveled by the photons at all shifts $|\Delta| < \Delta_{\rm peak}$. After each collision, the mean square number of mfps is $2\epsilon^{-2}$. Consequently, the total mean square movement is $\sim N\int_0^{\Delta_{\rm peak}} d\Delta\,\epsilon(2\epsilon^{-2}) \sim \tau^2$. That is, the photon diffuses $\sim \tau$ mfps, comparable to the number it moves in its largest single scatter. The order of this effect is clear in Figure 3. If the escape were primarily diffusive, then most of the photons would escape from $\tau \lesssim 1$. By contrast, we find that $\sim 1\%$ of the photons escape from regions of $\tau < 1$, $\sim 5\%$ from $\tau < 10$ and $\sim 14\%$ from $\tau < 100$. These values correspond to column densities of $N \sim 10^{14}\,\mathrm{cm}^{-2}$, $N \sim 10^{15}\,\mathrm{cm}^{-2}$, and $N \sim 10^{16}\,\mathrm{cm}^{-2}$. Thus, even if an optically thick Ly$\alpha$ cloud has an extended halo of relatively low density gas, the emission of Ly$\alpha$ comes primarily from the high $\tau$ region where the ionizing photons are absorbed. This validates the assumption, made in § 2, that the clouds can be considered "sharp-edged".

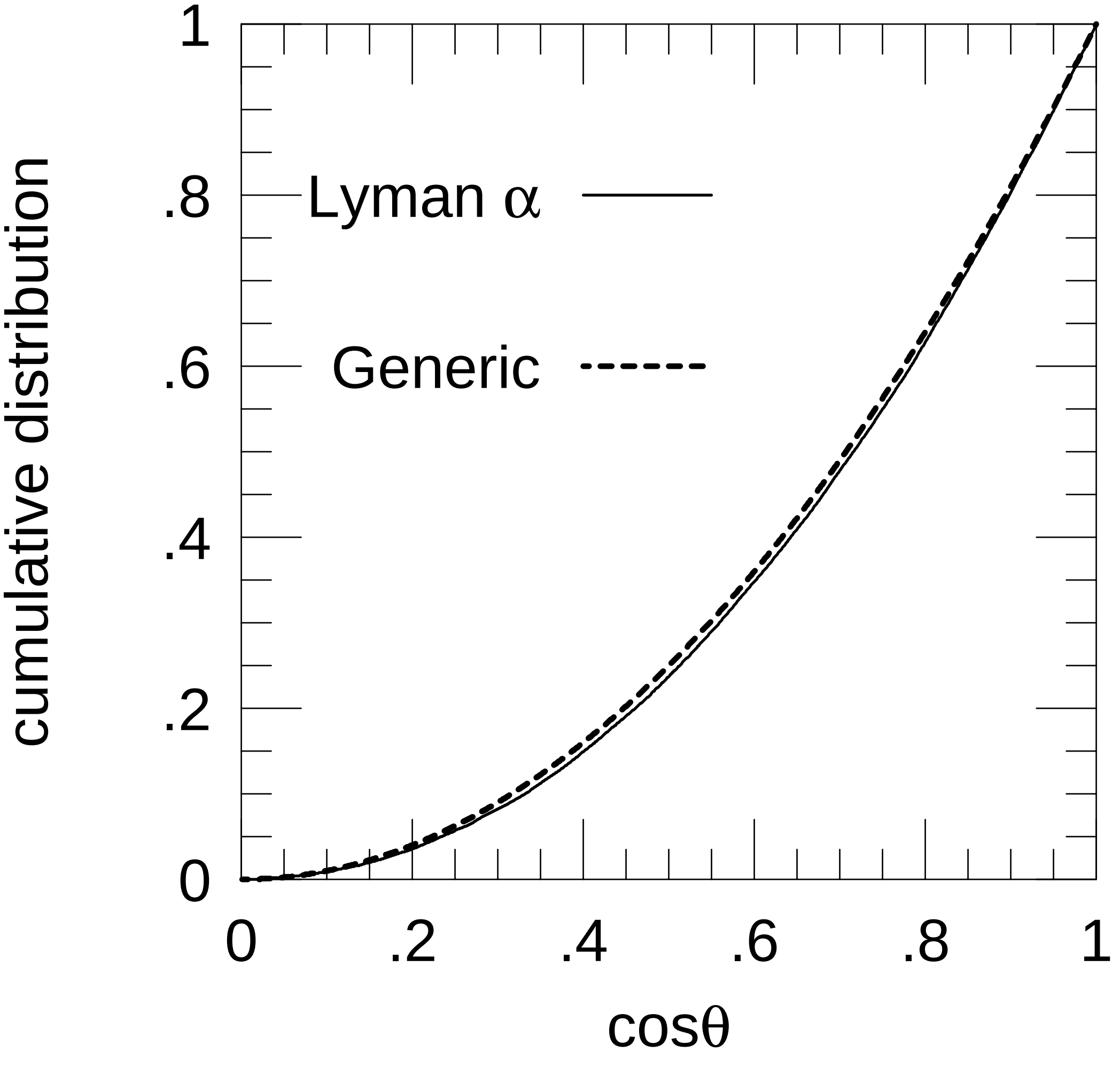

Figure 1. Cumulative distribution of emitted Ly$\alpha$ photons as a function of the cos of the angle $\theta$ relative to the normal of the surface (*solid curve*). The emitting cloud is assumed to have a slab geometry with a column $N = 10^{19}\,\mathrm{cm}^{-2}$. Shown for comparison is the distribution expected for optically thick diffusive emission (*dashed curve*). The fact that the two curves are almost identical means that the observed intensity of Ly$\alpha$ photons does not depend on the orientation of the cloud surface.

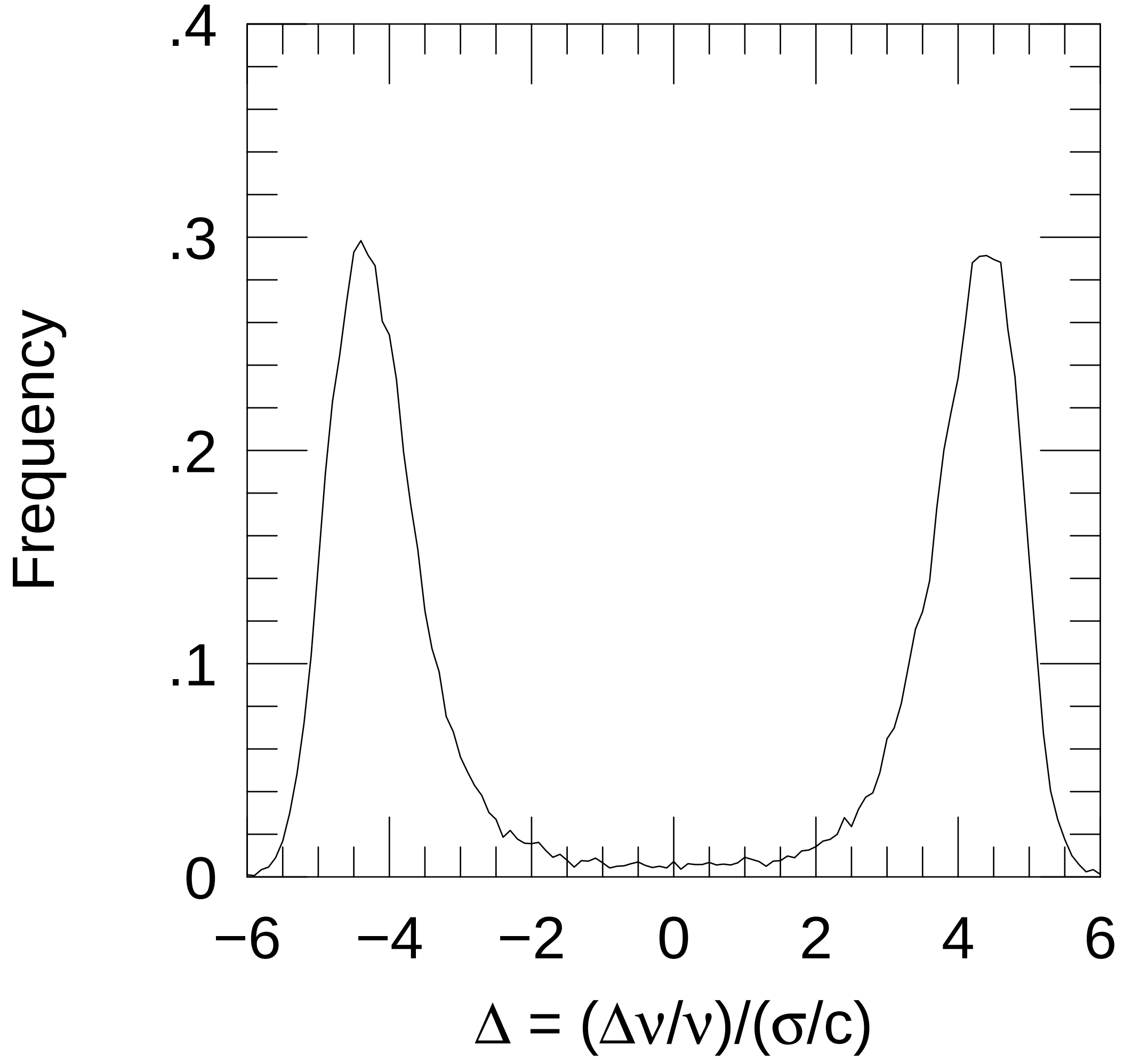

Figure 2. Distribution of photons as a function dimensionless Doppler shift, $\Delta$, of the Ly$\alpha$ photons emitted by an optically thick ($N = 10^{19}\,\mathrm{cm}^{-2}$) cloud.

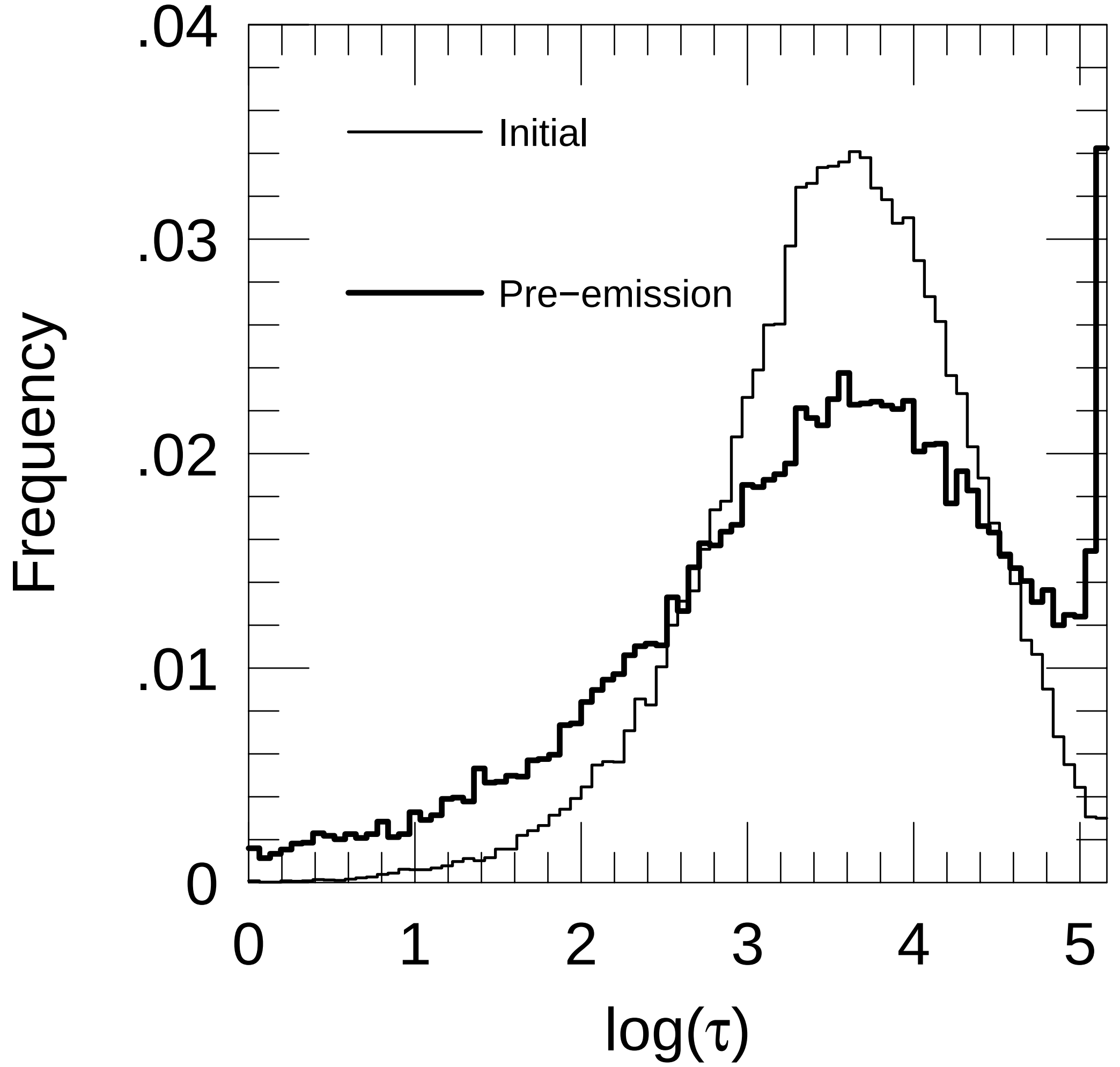

Figure 3. Initial (*solid histogram*) and pre-emission (*solid bold histogram*) distributions of photons in units of one Ly$\alpha$ optical depth. The distributions are broadly similar. The main difference is that photons tend to diffuse a factor $\sim 2$ closer to the edge before they escape. At $N = 10^{19}\,\mathrm{cm}^{-2}$, a fraction $\sim 87\%$ leave from the same side (left) that they entered. The remaining 13% escape from the far side, including $\sim 4\%$ that escape from the right-most bin which contains $\sim 2 \times 10^4$ optical depths.

Unlike Ly$\alpha$ photons created in star forming regions, the Ly$\alpha$ photons created by the ionizing background are virtually unaffected by dust. This is because the ionizing background penetrates only to a depth $N \sim 10^{18}\,\mathrm{cm}^{-2}$, so the Ly$\alpha$ photons have only to traverse this depth in order to escape. By contrast, the photons generated by hot stars deep within a damped Ly$\alpha$ system ($N \sim 10^{21}\,\mathrm{cm}^{-2}$) must travel through a column density $\sim 1000$ times larger.

Figure 4 shows the relative importance of various loss mechanisms — passage through cloud, direct capture to ground state, cascade through two-photon continuum, and extinction by dust — for slabs with column densities $N = 10^{17} - 10^{20}\,\mathrm{cm}^{-2}$. The highest line shows the fraction of ionizing photons which are initially captured by the cloud. The next three lines are the fractions which produce line photons, which produce Ly$\alpha$ photons, and which produce Ly$\alpha$ photons that escape before hitting a dust grain. At $N = 10^{20}\,\mathrm{cm}^{-2}$ virtually all ($> 98\%$) of the photons which escape come out the side that they entered. This means that for all $N > 10^{20}\,\mathrm{cm}^{-2}$, the results will be the same as for the final point. Thus, the emitted flux peaks at $N \sim 10^{19}\,\mathrm{cm}^{-2}$, but it is close to the peak for all column densities $N \gtrsim 10^{18.5}\,\mathrm{cm}^{-2}$.

## 4. Effects of Cloud Topology

In § 3, we modeled the cloud as having a slab-like shielding area with a single characteristic velocity dispersion. An alternative model for optically thick clouds is that they are composed of many cloudlets orbiting in a halo of hot, thermally unstable gas (e.g. Mo 1995). The observed velocity dispersion would then be due to the bulk motion of the individual cloudlets rather than the internal motions. Thus, the observed absorption line would be a composite of absorption from several individual cloudlets (the halo might contain many cloudlets, but only a few would intersect a given line of sight).

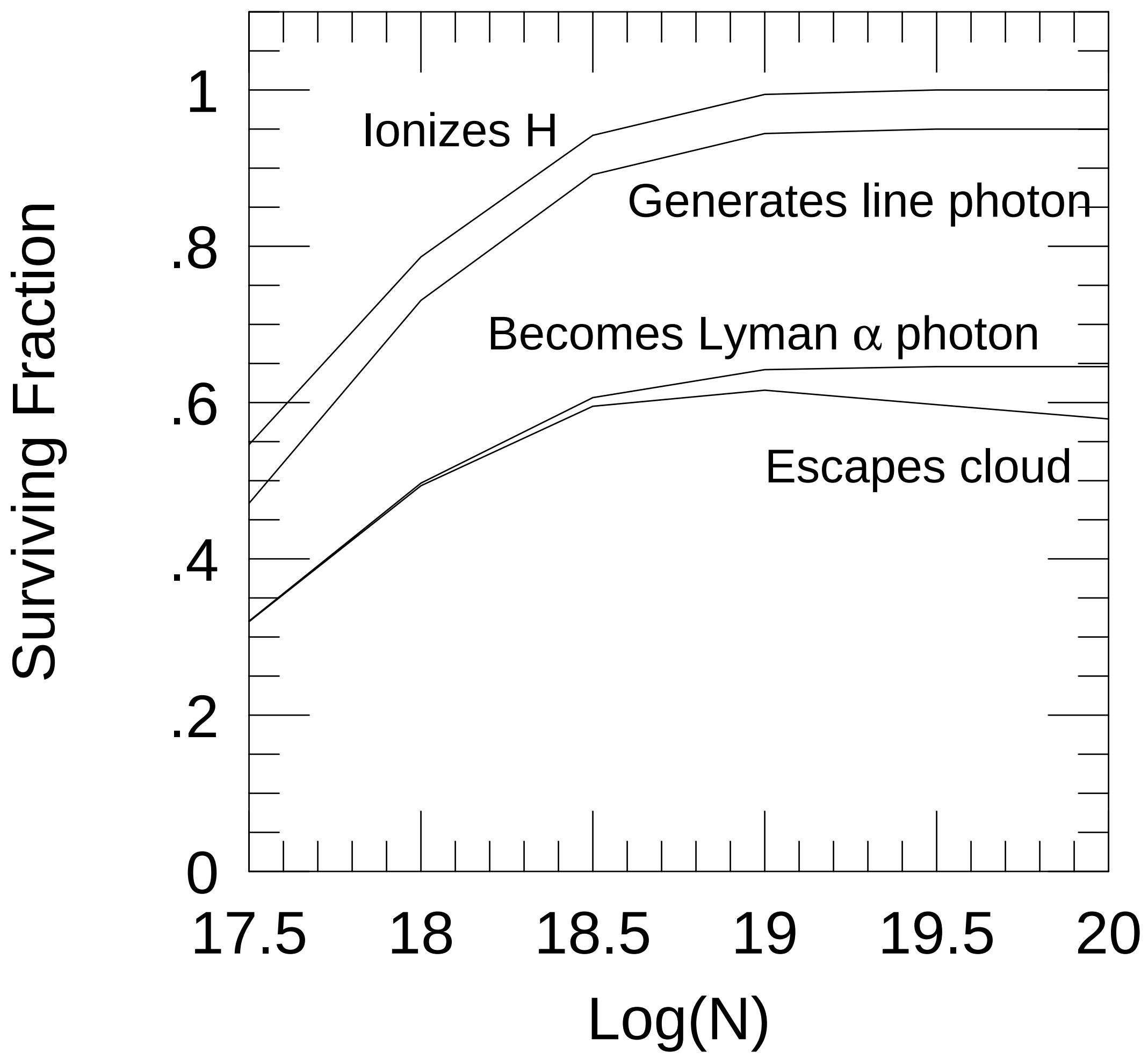

Figure 4. Fraction of ionizing photons which ultimately lead to Ly$\alpha$ emission as a function of cloud column density $N$. A slab geometry is assumed. The four curves shown represent the fraction of ionizing photons which lead to hydrogen ionizations, line photons, Ly$\alpha$ photons, and emitted Ly$\alpha$ photons.

Introducing this model has virtually no effect on the surface brightness of Ly$\alpha$ emission. It is still the case that essentially all ionizing radiation incident on the cloud will be absorbed. The Ly$\alpha$ yield from this absorption will be essentially unchanged. What will change is the profile of emitted radiation. Consider an observation of emission toward a quasar with a known system along the line of sight. To be specific, let the individual cloudlets have dispersions $\sigma \sim 10\,\mathrm{km\,s^{-1}}$ and be moving with respect to one another at $\sigma' \sim 70\,\mathrm{km\,s^{-1}}$. Suppose that the spatial resolution is good enough to resolve the individual cloudlets in the cloud system. What would one see directly on the line toward the quasar? First, one would see emission with a profile of the form shown in Figure 2, two sharp peaks separated by $\sim 8\sigma \sim 80\,\mathrm{km\,s^{-1}}$. This would come from the foremost cloudlet along this line of sight. There would be other cloudlets behind it (accounting for the observed dispersion in the absorption line), and these would emit similar double-peaked profiles (with different line centers). However, if any of these peaks fell between the two observed peaks from the front most cloudlet, these peaks would be absorbed. If any of the cloudlets had peaks outside this range, they would be transmitted through the front most cloud. Hence the overall spectrum would have a gap of $\sim 80\,\mathrm{km\,s^{-1}}$ surround by two (or possibly more) peaks.

If one could not resolve the individual cloudlet surfaces, either because of inadequate angular resolution or because of poor signal to noise, then one would see a superposition of such structures each with a different center. If the cloudlets were at high Mach number (large bulk velocities compared to internal dispersion), the emission profile would be fairly similar to the absorption profile.

## 5. Collisional Excitation

For the $N \sim 10^{19}\,\mathrm{cm^{-2}}$ clouds that we have been considering, cloud cooling is likely to be dominated by emission from collisionally excited HI (see e.g., figs. 1 and 2 of Katz, Weinberg, & Hernquist 1996). Here we show that for such clouds, $\sim 50\%$ of the energy of ionizing photons is emitted in the form Ly$\alpha$ photons. For simplicity of exposition, we initially consider that the ionizing spectrum is restricted to $\nu_0 < \nu < 4\nu_0$. We include photons with $\nu > 4\nu_0$ below and show that $\sim 50\%$ of their energy also emerges in Ly$\alpha$.

Consider an ionization by a photon with $\nu_0 < \nu < 4\nu_0$. The resulting electron has energy $h(\nu - \nu_0)$. The electron thermalizes and (assuming ionization equilibrium) is eventually recaptured when it has kinetic energy $h\nu_T \lesssim kT$. Hence, the original energy of the photon can be divided into two pieces: the energy (relative to the ground state) of the recombination electron, $h\nu_r$, and the thermal energy donated to the gas, $h\nu_g$. Explicitly,

$$\nu_r = \nu_0 + \nu_T, \qquad \nu_g = \nu - \nu_0 - \nu_T. \tag{5.1}$$

The final results depend only weakly on the mean kinetic energy at recombination. We henceforth adopt $\langle \nu_T \rangle = 0.1\nu_0$.

Our basic line of argument is that about half of each of these two pieces of energy is separately converted into Ly$\alpha$ photons. We focus first on the recombination piece. Recall that each ionization leads to $\sim 0.62$ Ly$\alpha$ photons. Hence the fraction of the recombination energy which emerges in Ly$\alpha$ is $\sim 0.62 \times 0.75\nu_0 / \langle \nu_r \rangle = 42\%$.

We now turn to the thermal energy, $h\nu_g$. We assume that 10% of this energy is emitted in Bremsstrahlung and that the rest goes into collisional excitation or ionization. We consider four classes of collisions, with final states 1) ionization, 2) $n = 2$, 3) $n \gg 2$, and 4) $n \gtrsim 3$.

If the atom is ionized, then the fraction of the energy converted into Ly$\alpha$ is the same as the recombination fraction computed above, 42%. Taking account of the Bremsstrahlung radiation reduces this to $0.9 \times 0.42 = 38\%$.

For all line excitations, 10% of the energy is lost to Bremsstrahlung, and 4% of the resulting Ly$\alpha$ photons are lost to dust (just as is true of Ly$\alpha$ photons generated by recombination). For $n = 2$, 25% go to $2S$ and 75% to $2P$. Hence the energy fraction in Ly$\alpha$ is $0.9 \times 0.96 \times 0.75 = 65\%$. For $n \gg 2$, excitations are mostly into high angular momentum states. Hence, the penultimate step of the cascade is nearly always $2P$, which always yields a Ly$\alpha$ photon. However, $\sim 25\%$ of the energy is lost in the cascade, so again the net fraction of energy in Ly$\alpha$ is 65%.

Finally, consider $n \gtrsim 3$. For $n = 3$, a fraction $5/32$ of the energy is lost to Balmer photons, and only $2/3$ of the resulting $n = 2$ states are $2P$. Hence, $9/16$ of the excitation energy goes to Ly$\alpha$ for a net fraction 49%. For $n > 3$, the fraction rises until it eventually reaches the $n \gg 2$ limit of 65%.

In fact, for plausible temperatures $T < 30,000$ K, there are very few ionizations, and most excitations are to $n = 2$. We therefore estimate that $\sim 60\%$ of $h\nu_g$ is emitted in Ly$\alpha$. With this estimate we find that the net fraction of ionization energy emitted in Ly$\alpha$ is

$$\eta_E \sim 0.50 + 0.1\left(1 - \frac{2\nu_0}{\langle\nu\rangle}\right), \tag{5.2}$$

where $h\langle\nu\rangle$ is the mean ionizing photon energy.

We now take account of energetic photons, with $\nu > 4\nu_0$. These require two types of correction, but both are small for the HM spectrum. First, if the photons are sufficiently energetic, then they will pass through the cloud without ionizing an atom. We identify the energy at which this is a significant effect by comparing the optical depth to HeII of a high energy photon with the optical depth to HI of a photon at the HeII ionization threshold, $4\nu_0$. The two are equal at $\nu \sim 16(r/4)^{4/11}\nu_0 \sim 40\nu_0$ where $r$ is the ratio of HeII to HI and where the evaluation is for $r = 50$. Thus, if a cloud is optically thick for $\nu < 4\nu_0$, it will also be optically thick for $\nu < 40\nu_0$. The precise value of this limit is not very important for the HM spectrum, since the absorption by the Ly$\alpha$ forest leaves relatively few photons above $4\nu_0$ in any case.

A second effect is that for each HeII ionization, a net of $\sim h\nu_0$ is lost during the process of HeII recombination. For the HM spectrum, only $\sim 5\%$ of the ionizing photons have $\nu > 4\nu_0$, while the mean frequency is $\langle\nu\rangle \sim 2.37\nu_0$. Hence, HeII recombination causes a loss of $\sim 2\%$ in the energy available for Ly$\alpha$ emission. We therefore ignore this effect.

If the cloud temperature is sufficiently low, then free-free and free-bound cooling may compete with line cooling, but this should only happen if the thermal energy input from the ionizing background is very low, in which case the Ly$\alpha$ emission from recombination would dominate over that from excitation anyway. Line cooling also becomes inefficient at high temperatures, but a hot cloud would be collisionally ionized and unlikely to be an optically thick system in the first place. Thus the conclusion that $\sim 50\%$ of the incident energy of the ionizing background emerges in the form of Ly$\alpha$ photons should be fairly robust.

We have assumed that the only energy input to the cloud is from photoionization and that other heating mechanisms, such as shocks dissipating internal motions, are unimportant. If other mechanisms do heat the cloud, some of this energy will also emerge in line cooling, enhancing the Ly$\alpha$ emission.

## 6. Observing the Forest

### 6.1. Signal and Noise

Combining the above argument for the total emission rate with the discussion in § 2 of the observed photon flux, we find

$$\begin{aligned}\Phi_{\rm obs} &= (1+z)^{-3}\eta_E \frac{\langle\nu\rangle}{(3/4)\nu_0}\int\limits_{\nu_0}^{\infty} d\nu\phi(\nu) \\ &= (1+z)^{-3}\frac{2}{3}\left(\frac{\eta_E}{0.5}\right)\frac{\langle\nu\rangle}{\nu_0}\int\limits_{\nu_0}^{\infty} d\nu\phi(\nu).\end{aligned} \tag{6.1}$$

Here $\eta_E$ is the fraction of energy absorbed from the ionizing background that is emitted in the form of Ly$\alpha$ photons. As argued in § 5, one expects $\eta_E \sim 0.5$, but given the range of plausible physical conditions, $\eta_E$ could be up to 20% higher or lower than this value.

Suppose that a Ly$\alpha$ cloud of angular extent $\Delta\Omega$ is observed for a time $T$ by a telescope of aperture $(\pi/4)D^2$ and efficiency $f$. The total number of photons collected will then be $n_s = (\pi/4)\zeta f D^2 T\Delta\Omega\Phi_{\rm obs}$, where $\zeta \sim 80\%$ is the fraction of photons penetrating the Earth's atmosphere. To determine the number of noise photons, $n_n$, we assume (based on the profile calculation in § 3) that the effective width of the double-peaked emission line is $[\int dv[g(v)]^2]^{-1} = 4\pi^{1/2}\sigma$, where $\sigma$ is the 1-dimensional cloud velocity dispersion and $g(v)$ is the profile shape. We assume that the spectrum is taken in the $B$ band with adequate resolution to sample the velocity structure of the line and with sky of $B = 22.2$ and hence a sky flux of $\phi_{\rm sky} = 1.8 \times 10^{-2}{\rm photons\,m^{-2}\,s^{-1}\,A^{-1}\,arcsec^{-2}}$. The number of sky photons is then $n_n = (\pi/4)D^2 fT\Delta\Omega\phi_{\rm sky}\lambda/c\int dv[g(v)]^2$, where $\lambda$ is the observed wavelength. Hence, the signal-to-noise ratio for an optically thick cloud is,

$$\frac{\rm S}{\rm N} = \zeta\frac{\Phi_{\rm obs}}{(1+z)^{1/2}}\left[\frac{\pi^{1/2}D^2 fT\Delta\Omega}{16(\sigma/c)\phi_{\rm sky}\lambda_{{\rm Ly}\alpha}}\right]^{1/2}$$
$$\sim 7.5\left(\frac{1+z}{3.2}\right)^{-3.5}\left(\frac{f}{0.25}\right)^{1/2}\left(\frac{D}{10\,{\rm m}}\right)\left(\frac{T}{20\,{\rm hour}}\right)^{1/2}\left(\frac{\Delta\Omega}{10\,{\rm arcsec}^2}\right)^{1/2}\left(\frac{\sigma}{35\,{\rm km\,s^{-1}}}\right)^{-1/2}. \quad (6.2)$$

Note that the evaluation is for a twin-peaked profile characteristic of a homogeneous cloud. For a Gaussian profile (which might be generated by an ensemble of small clouds traveling at high Mach number – see § 5) the same $S/N$ is achieved for twice the dispersion, i.e. $\sigma = 70\,{\rm km\,s^{-1}}$.

In making our numerical evaluation, we have adopted the values $\langle\nu\rangle = 2.37\nu_0$ and $\int_{\nu_0}^{\infty} d\nu\phi(\nu) = 5.3\times 10^4\,{\rm cm^{-2}\,s^{-1}\,sr^{-1}}$ based on the HM spectrum. One would obtain the same values from a background with energy spectrum $J(\nu) = 6.1 \times 10^{-22}(\nu/\nu_0)^{-1.73}\,{\rm erg\,cm^{-2}\,s^{-1}\,sr^{-1}\,Hz^{-1}}$ . Uncertainties in the amplitude of the ionizing background introduce a factor $\sim 2$ uncertainty in the predicted $S/N$. In particular, if star-forming galaxies or dust-obscured quasars contribute to the background, the signal could be significantly higher. From the form of equation (6.2), it is clear that the prospect of detecting Ly$\alpha$ emission declines rapidly with redshift, and therefore that observations should first be attempted at the minimum redshift where there is good system throughput.

## 6.2. Bare Detection

As indicated by equation (6.2), even to detect Ly$\alpha$ emission requires a long integration of an optically thick cloud with a high-throughput spectrograph on a large telescope. For an initial experiment, it is essential therefore to choose a system which is known to exist and to be optically thick from its absorption of quasar light. At $N \sim 10^{17}\,\mathrm{cm}^{-2}$, $\sim L_*$ galaxies observed at $z \sim 0.5$–$1.5$ produce absorption out to $\sim 40h^{-1}$ kpc spherical radius (Steidel & Dickinson 1995), where $h$ is the Hubble parameter in units of $100\,\mathrm{km\,s^{-1}\,Mpc^{-1}}$. Since $1'' \approx 4h^{-1}$ kpc (at z=2.2), this corresponds to a projected diameter $\sim 20''$. Simulations (Katz et al. 1995) suggest that at $N \sim 10^{19}\,\mathrm{cm}^{-2}$ the clouds are smaller by a factor $\sim 2$. Thus a long $1''$ slit might cover $\sim 10\,\mathrm{arcsec}^2$ of a cloud.

For an initial detection one should therefore place as wide a slit as possible over a quasar with a known system at $z \sim 2.2$ along the line of sight. The cloud will itself absorb most of the quasar light near Ly$\alpha$, but to minimize contamination, one should choose the quasar to be as *faint* as possible. If the system does extend over many square arcsec, it should contain emitting regions well away from the quasar.

To maximize the S/N, one would like to resolve the lines. In fact, just to match the expected width of the lines $\sigma \sim 35\,\mathrm{km\,s^{-1}}$ (as assumed in making the S/N estimates in § 3) requires a resolution $\sim 8,000$. Still higher resolution would be better, but it is probably inconsistent with high throughput at the present time.

A measurement of the Ly$\alpha$ surface brightness from an optically thick cloud would yield a direct estimate of the integrated energy flux of the ionizing background through equation (6.1). As shown in § 3, the emitted surface brightness is virtually independent of cloud geometry, so this estimate should be robust. The principle theoretical uncertainty is the possible contribution of heating sources other than photoionization; if other sources are important, then the estimate of the background energy density becomes an upper limit. This issue can be investigated empirically by comparing results from a number of systems. If they consistently yield the same surface brightness, then it is likely that internal heating is unimportant and one has estimated the background energy density. If they yield a range of values, then one has an upper limit on the background and some information about the magnitude and range of internal heating sources.

## 6.3. Mapping High-Redshift Structure

At present the Ly$\alpha$ forest is generally studied in only one dimension: along isolated lines of sight to quasars. These lines typically contain of order 2–3 Lyman limit system per unit redshift. If the clouds are indeed $5''$ in radius, then a $1.''5 \times 1000''$ slit should contain 500 systems per unit redshift. Assuming 2 pixels per resolution element, a $2048^2$ CCD could cover a redshift range $\Delta z \sim 0.4$ near $z \sim 2.2$. Hence, one could hope to find 200 new systems as part of the same observations used to detect a known absorption system in emission. This would provide a good statistical measure of the number density and sizes of such systems.

In addition, once the possibility of finding such field systems were demonstrated, one could initiate searches in regions of the sky not associated with quasars. By studying the distribution of Ly$\alpha$ clouds in two dimensions (redshift and position along slit), one could obtain important constraints on their correlations and geometry, and ultimately on their physical nature. A more ambitious project would be to take spectra along adjacent slit positions and so build up a 3-dimensional picture of the Ly$\alpha$ forest.

Observations like these could also "hitchhike" on programs requiring long duration, high throughput spectroscopic exposures of the appropriate resolution. They could easily go hand-in-hand with searches for primeval galaxies. Indeed, it is quite possible that the first large population of objects detected in such searches will not be galaxies emitting the energy produced by their own stars but hydrogen gas clouds fluorescing in the UV background.

**Acknowledgements**: We thank Jordi Miralda-Escudé for pointing out the important role of line cooling emission and for helpful discussions about possible topologies of the emitting regions. We thank James Lowenthal and Craig Hogan for helpful discussions and references to earlier work. We thank Piero Madau for providing the Haardt & Madau (1995) spectrum in a convenient numerical form, prior to publication. Work by A.G. was supported in part by grant AST 94-20746 from the NSF and by grant NAG5-3111 from NASA. Work by D.W. was supported in part by NASA grants NAG5-2882 and NAG5-3111.